\documentclass[runningheads,fleqn]{llncs}
\usepackage[T1]{fontenc}
\usepackage{graphicx}
\usepackage{tikz}
\usepackage{listings}
\usepackage[hidelinks]{hyperref}
\usetikzlibrary{math}

\lstdefinestyle{mystyle}{
  language=C,
  breaklines=true, 
  basicstyle=\ttfamily\footnotesize
}
\newcommand{\sessions}{MPI Sessions}
\newcommand{\WORLD}{\texttt{MPI\_COMM\_WORLD}}
\newcommand{\world}{\texttt{comm\_world}}
\newcommand{\SessionInit}{\texttt{MPI\_Session\_init}}

\begin{document}

\title{Implementing True MPI Sessions and Evaluating MPI Initialization Scalability}

% \author{Hui Zhou\inst{1} \and
% Ken Raffenetti\inst{1} \and
% Mike Wilkins\inst{1} \and
% Yanfei Guo\inst{1} \and
% Rajeev Thakur\inst{1}}
\author{Hui Zhou\inst{1}\orcidID{0000-0002-4422-2911} \and
Kenneth Raffenetti\inst{1}\orcidID{0009-0003-4705-2713} \and
Yanfei Guo\inst{1}\orcidID{0000-0002-3731-5423} \and
Michael Wilkins\inst{1}\orcidID{0000-0003-0806-1599} \and
Rajeev Thakur\inst{1}\orcidID{0000-0002-5532-3048}}
\authorrunning{H. Zhou et al.}
\titlerunning{Implementing True MPI Sessions}

\institute{Argonne National Laboratory, Lemont IL 60439, USA \\
\email{hzhou321,raffenet,wilkins,yguo,thakur@anl.gov}\\
%\url{http://www.anl.gov}
}
\maketitle              % typeset the header of the contribution
\begin{abstract}

Sessions is one of the major features introduced in the MPI-4 standard. It offers an alternative to the traditional world communicator model by allowing applications to construct communicators from process sets, thereby eliminating the dependency on \WORLD. The Sessions model was proposed as a more scalable solution for exascale systems, where \WORLD\ was viewed as a potential scalability bottleneck. However, supporting Sessions is a significant challenge for established codebases like MPICH due to the deep integration of the world model in traditional MPI implementations.
Although MPICH added support for the MPI-4 standard upon its release, it still internally relied on a global world communicator. 
This approach enabled applications written using the Sessions model to function, but it did not fulfill the full design intent of Sessions, which meant to decouple MPI from \WORLD.
We describe MPICH's effort to support  ``true'' MPI Sessions, including a major internal refactoring. We describe the architectural changes required to support true Sessions and evaluate the resulting implementation’s scalability. Our results demonstrate that true Sessions can offer significant scalability benefits by adopting explicit hierarchical designs. 

%\keywords{Message Passing Interface \and MPI Sessions \and MPICH}
\end{abstract}

\section{Introduction}
The MPI Forum released Version 4 of the Message Passing Interface (MPI) Standard in 2021 \cite{mpi40}, introducing several major features, most notably \sessions ~\cite{holmes2016mpi}.
The \sessions\ model offers an alternative to the traditional MPI paradigm by enabling applications to construct communicators from locally defined process sets, rather than relying on the default global communicator, \WORLD, and deriving all others from it.

The primary motivation behind \sessions\ was to remove the dependency on \WORLD, which was anticipated to be a major scalability barrier to exascale computing. However, concerns over the scalability of \WORLD\ appear to have been overstated. Since the petascale era, MPI has proven its ability to scale to over a million processes \cite{balaji2009mpi}, while the anticipated billion-process scale has yet to materialize. In practice, exascale computing has been successfully achieved, largely due to advancements like massively parallel accelerators (e.g., GPUs), without the need to eliminate \WORLD. In fact, a primary objective for exascale facilities has been to achieve a 50× performance improvement over previous systems without necessitating substantial code modifications in existing applications \cite{atchley2023frontier}. Empirical evidence suggests that \WORLD\ has not posed a fundamental limitation to scalability.

Another motivation for introducing \sessions\ was to address MPI’s limited support for dynamic behavior. Various software components can initialize separate \sessions\ independently without coordination. This includes non-overlapping Sessions, effectively enabling MPI to be initialized and finalized multiple times within a single application. Such capability is particularly beneficial for workflow-oriented applications that orchestrate multiple independent tasks, each potentially using MPI \cite{wozniak2019mpi}.
Second, the \sessions\ functionality includes an info argument in \texttt{MPI\_Session\_init}, allowing customization of each session with parameters that are typically global, such as MPI thread levels.
Third, the dynamic nature of \sessions\ enables innovative solutions for fault tolerance and resource management, including the ability to shrink or grow the number of participating processes during an application's execution.

% Moreover, MPI already includes support for dynamic processes, which can address several of the same needs, particularly in areas such as workflows and resource malleability.
% potentially consider it in discussion

The inclusion of \sessions\ in MPI 4.0 has reignited interest and spurred research into addressing some long-standing challenges. Active areas of exploration include resource management \cite{dosanjh2021implementation,fecht2022emulation}, fault tolerance \cite{rocco2023fault}, and job malleability \cite{suarez2024deep}. However, to date, there have been limited evaluations of the scalability of \sessions.

% Introduce MPICH, justify the motivation to support true MPI sessions.

MPICH \cite{gropp1996high} has played a critical role in MPI’s evolution. It was the first widely adopted implementation and served as a prototyping platform during the development of MPI-1. During MPI-2, MPICH supported the standardization process through early implementations of features like ROMIO for MPI-IO \cite{thakur1999data}. Although this implementation-driven standardization model weakened during MPI-3, MPICH still became the first full implementation of the MPI-3 standard. Continuing this tradition, MPICH released full support for MPI-4—including \sessions—in the same month the standard was published.

However, MPICH’s initial \sessions\ implementation relies on an internal world communicator behind the scenes. Therefore, while MPICH provides the MPI Sessions interface, it is limited to standard-defined process sets, namely, \texttt{mpi://WORLD} and \texttt{mpi://SELF}. Such limited support is permitted by the MPI specification but is not in line with the ``true'' spirit of the \sessions\ proposal that saw a world communicator as optional. Developers who wished to explore the possibilities of \sessions\ either had to build their own solution or look to other MPI libraries with more full-featured support.

Implementing a true \sessions\ model in MPICH presents significant challenges for two main reasons. First, MPI was originally designed around \WORLD, which abstracted away the need for applications to manage process identities beyond simple integer ranks. This design choice simplified the application interface and allowed implementations internal flexibility. MPICH, in particular, uses a device-dependent model with internal, implementation-specific process-addressing schemes. However, a true Sessions model demands a device-independent, standardized process-addressing mechanism, which requires major architectural changes in MPICH.

Second, MPICH's design philosophy clearly separates performance-critical runtime paths from less performance-critical initialization stages. It avoids dynamic mechanisms in favor of a structured, two-phase model consisting of an initialization phase followed by a runtime phase. The true \sessions\ model breaks the long standing assumption that a single, global initialization phase always occurs---an assumption that MPICH relies on for efficient global setup. This shift necessitates a significant redesign of MPICH’s initialization and resource-management infrastructure.

% motivation for this work
However, there are also strong motivations for MPICH to support true \sessions. Beyond enabling ongoing research efforts and accommodating next-generation HPC applications that embrace more dynamic designs, implementing a device-independent infrastructure for process sets and MPI groups has the potential to simplify MPICH's device layer and promote more sustainable development and maintenance. Furthermore, the effort to support true \sessions\ aligns with MPICH’s broader goal of strengthening support for dynamic processes, which already serve a sizable user base.

%MW: I feel like the introduction ends awkwardly here. I expect an introduction to also summarize the contributions of the paper and tease the results.
To this end, we recently performed a substantial code restructuring in MPICH to enable support for true \sessions. Following this work, we evaluated the scalability of initialization across three MPI implementations: (1) the newly refactored MPICH with true \sessions\ support, (2) MPICH version 4.3.0, which still relies on an internal world communicator, and (3) Open MPI version 5.0.7, which also includes support for true \sessions. These evaluations were conducted on Aurora, the new exascale system at the Argonne Leadership Computing Facility (ALCF).
Our findings indicate that \sessions\ provide comparable initialization scalability across all three implementations on this system. Notably, true \sessions\ enable the construction of a sparsely connected world communication pattern, which may reduce resource usage during initialization, albeit modestly.

% Introduce the content of the this paper
% MW: If we need space, this paragraph can be entirely removed. I am generally adverse to including outlines like this in my papers. You can elude to the structure of the paper by referencing sections in previous paragraphs.
% The rest of the paper is organized as follows. Section~2 provides essential background, including an overview of MPICH's architecture, the Process Management Interface (PMI), and key considerations for efficient MPI startup. Section~3 details MPICH’s implementation of \sessions, beginning with earlier work that supported \sessions\ via an internal world communicator, followed by a series of recent refactoring work that enable support for true \sessions. In Section~4, we evaluate the scalability of \sessions\ by comparing them with the traditional \WORLD\ model using Open MPI version~5, MPICH version~4.3.0 (which supports \sessions\ via an internal \WORLD), and the latest MPICH development branch, which supports true \sessions. Section~5 discusses the lessons learned during the implementation process. Finally, Section~6 concludes the paper.

\section{Background}
% need sentences 
We begin with some essential background, including an overview of MPICH's architecture, the Process Management Interface (PMI), and key considerations for efficient MPI startup.

\subsection{MPICH Architecture}

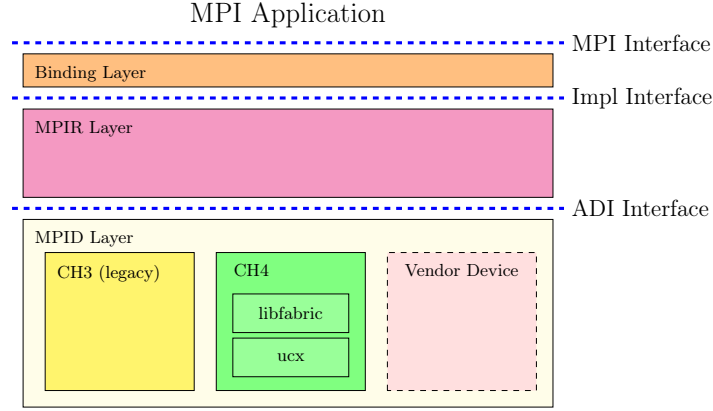
\begin{figure}[t]
\centering
\resizebox{0.8\textwidth}{!}{
    \tikzstyle{box}=[draw, rectangle, fill=white, minimum height=0.25cm]
\tikzstyle{interface}=[draw=blue, ultra thick, dashed]

\begin{tikzpicture}

\node at (5, -0.5) {\Large MPI Application};

\draw[interface] (0, -1) -- (10, -1) node[right=0.2] {\large MPI Interface};

\draw[fill=orange!50!white] (0.2, -2+0.2) rectangle (10-0.2, -1-0.2);
\node[anchor=north west] at (0.3, -1-0.3) {Binding Layer};

\draw[interface] (0, -2) -- (10, -2) node[right=0.2] {\large Impl Interface};

\draw[fill=magenta!50!white] (0.2, -4+0.2) rectangle (10-0.2, -2-0.2);
\node[anchor=north west] at (0.3, -2-0.3) {MPIR Layer};

\draw[interface] (0, -4) -- (10, -4) node[right=0.2] {\large{ADI Interface}};

\draw[fill=yellow!10!white] (0.2, -7.6) rectangle (10-0.2, -4-0.2);
\node[anchor=north west] at (0.3, -4-0.3) {MPID Layer};

\draw[fill=yellow!70!white] (0.6, -7.5+0.2) rectangle (3.5-0.2, -4.6-0.2);
\node[anchor=north west] at (0.7, -4.6-0.3) {CH3 (legacy)};
\draw[fill=green!50!white] (3.5+0.2, -7.5+0.2) rectangle (6.6-0.2, -4.6-0.2);
\node[anchor=north west] at (3.6+0.3, -4.6-0.3) {CH4};
\node[draw, fill=green!40!white, minimum width=2.1cm, minimum height=0.7cm] at (5.05, -5.9) {libfabric};
\node[draw, fill=green!40!white, minimum width=2.1cm, minimum height=0.7cm] at (5.05, -6.7) {ucx};

\draw[fill=pink!50!white, dashed ] (6.6+0.2, -7.5+0.2) rectangle (9.7-0.2, -4.6-0.2);
\node[anchor=north west] at (6.7+0.3, -4.6-0.3) {Vendor Device};

\end{tikzpicture}
}    
\caption{MPICH architecture. The binding layer handles parameter checking and converts MPI object handles to internal structure pointers. The MPIR layer provides device-independent utilities, while the device layer implements hardware-specific functionality. MPICH maintains CH3 and CH4 as reference device implementations. Vendors may adopt CH3/CH4 or implement their own devices that conform to the ADI interface.}
\label{fig:mpich-arch}
\end{figure}

MPICH adopts an architectural design that facilitates vendor adoption, as illustrated in Figure~\ref{fig:mpich-arch}. The key abstraction is the ADI (Abstract Device Interface), which separates the codebase into two primary layers: the MPIR (MPI Runtime) layer and the MPID (MPI Device) layer. An additional binding layer sits above the MPIR layer, handling parameter validation and conversions~\cite{zhou2024generating}.

The architectural design allows the MPICH project to focus on MPI semantics, implementation strategies, and high-level algorithms, while enabling MPI vendors to concentrate on hardware-specific performance optimizations.
Broadly, the MPIR layer is responsible for hardware-independent functionality, while the MPID layer handles hardware-specific implementations. However, this separation is not absolute. In most communication routines, after the binding layer completes parameter checks and conversions, control is passed quickly to the corresponding ADI routine in the device layer. This design allows the device layer to take the full control of performance-critical code paths, ensuring no optimization opportunities are lost~\cite{guo2025preparing}.

\iffalse
%MW: I think the paragraph and/or the figure below could be eliminated to save space.
For fast-path functions such as \texttt{MPI\_Send} and \texttt{MPI\_Put}, the MPIR layer introduces minimal overhead, enabling near-metal, low-latency performance \cite{guo2025preparing}. A typical call path for \texttt{MPI\_Send} is:
%\begin{displaymath}
    \texttt{MPI\_Send} $\rightarrow$ \texttt{MPID\_Send}.
%\end{displaymath}
For collective operations, the MPIR layer provides hardware-independent implementations of collective algorithms. However, the call path routes through the device layer early, allowing vendors to either override the operation with hardware-accelerated implementations or fallback to the MPIR algorithms, depending on performance considerations. For example, the call path for \texttt{MPI\_Bcast} might look like:
\begin{displaymath}
    \texttt{MPI\_Bcast} \rightarrow \texttt{MPID\_Bcast} \rightarrow \texttt{MPIR\_Bcast\_impl} %MW: I would drop "_impl", even though I know its part of the real function name.
\end{displaymath}
or: \hskip .17in
%\begin{displaymath}
    \texttt{MPI\_Bcast} $\rightarrow$ \texttt{MPID\_Bcast} $\rightarrow$ \texttt{MPIDI\_Bcast\_hardware}
%\end{displaymath}
\fi

The ADI design introduces some complexity, particularly during the initialization stage. Subcomponents within both the MPIR and device layers require initialization and may depend on one another, requiring a carefully coordinated initialization order. 
As a result, MPICH adopts a roughly three-stage initialization process, illustrated in the following pseudocode:
\begin{lstlisting}
    void MPIR_Init() {
        MPIR_Pre_Init();  /* dev-independent init */
        MPID_Init();      /* dev-dependent init */
        MPIR_Post_Init(); /* dev-dependent MPIR-layer init */
    }
\end{lstlisting}

However, this initialization process mixes local and collective steps. A local initialization step does not involve communication or synchronization with other processes. In contrast, a collective initialization step involves coordination across processes, often requiring data exchange and execution barriers. %MW: the first sentence in this paragraph is awkward. I suggest "The current initialization process freely mixes local and collective steps." HZ: done.
In the pseudocode above, \texttt{MPIR\_Pre\_Init} is entirely local, but both \texttt{MPID\_Init} and \texttt{MPIR\_Post\_Init} may contain a mix of local and collective components.
This interleaving poses a major challenge in the context of \sessions, because \texttt{MPI\_Session\_Init} is defined by the MPI standard as a strictly local-only routine. 
Isolating local-only initialization from this tightly coupled structure requires significant restructuring across layers (vertically) and components (horizontally).
Furthermore, \sessions\ do not assume the presence of a global, world-level collective initialization stage. As a result, any initialization logic that previously relied on this assumption must be redesigned. Collective components must now operate under weaker assumptions, increasing complexity and implementation effort.

% -------------------------------------------------------------------
%MW: This section is interesting to me to learn the history of PMI, but I think it can be shortened significantly.
% AFAIK, the detail and differences between PMI version is not necessary to understand this paper.
\subsection{Process Management Interface (PMI)} \label{PMI}
The MPI specification omits an important detail: how processes are launched and how they exchange wire-up information to bootstrap a communicator. To fill this gap, MPICH introduced the Process Management Interface (PMI) \cite{balaji2010pmi}. The original version, known as PMI-1, is distributed with MPICH as a header file, \textit{pmi.h}, and is widely adopted by most MPI implementations and process managers, including Slurm \cite{yoo2003slurm} and Flux \cite{ahn2014flux}. MPICH also provides its own PMI implementation in its upstream releases, including a job launcher, hydra \cite{balaji2010pmi}.

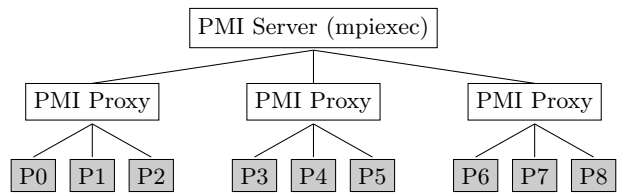
\begin{figure}[t]
\centering
\resizebox{0.7\textwidth}{!}{
    \tikzstyle{pmibox}=[draw, rectangle, fill=white, minimum height=0.25cm]
\tikzstyle{procbox}=[draw, rectangle, fill=gray!40, minimum height=0.25cm]

\begin{tikzpicture}

\node[pmibox] at (5, 0) {PMI Server (mpiexec)};

\node[pmibox] at (2, -1) {PMI Proxy};
\node[pmibox] at (5, -1) {PMI Proxy};
\node[pmibox] at (8, -1) {PMI Proxy};

\node[procbox] at (1.2,-2) {P0};
\node[procbox] at (2,-2) {P1};
\node[procbox] at (2.8,-2) {P2};

\node[procbox] at (4.2,-2) {P3};
\node[procbox] at (5,-2) {P4};
\node[procbox] at (5.8,-2) {P5};

\node[procbox] at (7.2,-2) {P6};
\node[procbox] at (8,-2) {P7};
\node[procbox] at (8.8,-2) {P8};

\draw (5, -0.3) -- (2, -0.75);
\draw (5, -0.3) -- (5, -0.75);
\draw (5, -0.3) -- (8, -0.75);

\draw (2, -1.3) -- (1.2, -1.75);
\draw (2, -1.3) -- (2.0, -1.75);
\draw (2, -1.3) -- (2.8, -1.75);

\draw (5, -1.3) -- (4.2, -1.75);
\draw (5, -1.3) -- (5.0, -1.75);
\draw (5, -1.3) -- (5.8, -1.75);

\draw (8, -1.3) -- (7.2, -1.75);
\draw (8, -1.3) -- (8.0, -1.75);
\draw (8, -1.3) -- (8.8, -1.75);
\end{tikzpicture}
}    
\caption{A common process launch mechanism in MPI. The PMI server (e.g., \textit{mpiexec}) launches PMI proxies (usually one per compute node), and each proxy launches the MPI processes for its node.}
\label{fig:pmi-launch}
\end{figure}

PMI does not define the process launch mechanism itself. For illustrative purposes, we refer to a common setup shown in Figure~\ref{fig:pmi-launch}. In this example, \texttt{mpiexec} is used to launch 9 MPI processes across 3 nodes. \texttt{mpiexec} acts as the PMI server, launching one PMI proxy per node, which in turn launches the MPI processes on that node. The PMI server maintains connections with the PMI proxies, typically using TCP, while each PMI proxy communicates with its local MPI processes using Unix pipes. Together, the PMI server, proxies, and MPI processes form a spanning tree that enables data exchange among all MPI processes.

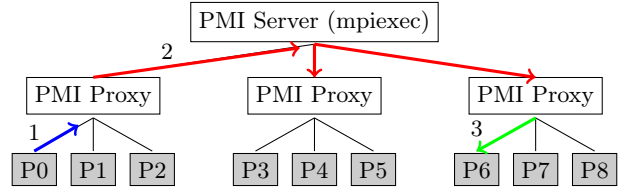
\begin{figure}[t]
\centering
\resizebox{0.7\textwidth}{!}{
    \tikzstyle{pmibox}=[draw, rectangle, fill=white, minimum height=0.25cm]
\tikzstyle{procbox}=[draw, rectangle, fill=gray!40, minimum height=0.25cm]

\begin{tikzpicture}

\node[pmibox] at (5, 0) {PMI Server (mpiexec)};

\node[pmibox] at (2, -1) {PMI Proxy};
\node[pmibox] at (5, -1) {PMI Proxy};
\node[pmibox] at (8, -1) {PMI Proxy};

\node[procbox] at (1.2,-2) {P0};
\node[procbox] at (2,-2) {P1};
\node[procbox] at (2.8,-2) {P2};

\node[procbox] at (4.2,-2) {P3};
\node[procbox] at (5,-2) {P4};
\node[procbox] at (5.8,-2) {P5};

\node[procbox] at (7.2,-2) {P6};
\node[procbox] at (8,-2) {P7};
\node[procbox] at (8.8,-2) {P8};

\draw (5, -0.3) -- (2, -0.75);
\draw (5, -0.3) -- (5, -0.75);
\draw (5, -0.3) -- (8, -0.75);

\draw (2, -1.3) -- (1.2, -1.75);
\draw (2, -1.3) -- (2.0, -1.75);
\draw (2, -1.3) -- (2.8, -1.75);

\draw (5, -1.3) -- (4.2, -1.75);
\draw (5, -1.3) -- (5.0, -1.75);
\draw (5, -1.3) -- (5.8, -1.75);

\draw (8, -1.3) -- (7.2, -1.75);
\draw (8, -1.3) -- (8.0, -1.75);
\draw (8, -1.3) -- (8.8, -1.75);

\draw[<-, very thick, draw=blue] (1.8, -1.4) -- (1.2, -1.75); \node at (1.2, -1.5) {1};
\draw[<-, very thick, draw=red]  (4.8, -0.35) -- (2, -0.75); \node at (3.0, -0.4) {2};
\draw[->, very thick, draw=red] (5, -0.3) -- (5, -0.75);
\draw[->, very thick, draw=red] (5, -0.3) -- (8, -0.75);

\draw[->, very thick, draw=green] (8, -1.3) -- (7.2, -1.75); \node at (7.2, -1.45) {3};
\end{tikzpicture}
}    
\caption{An example of data exchange using PMI. (1) P0 use \texttt{PMI\_Put} to send data to its PMI proxy. (2) All MPI processes call \texttt{PMI\_Barrier}, during which proxies synchronize local data to the PMI server, and the PMI server propagates the data to other proxies. (3) P6 calls \texttt{PMI\_Get} to retrieve the data from its local proxy.}
\label{fig:pmi-kvs}
\end{figure}

Figure~\ref{fig:pmi-kvs} illustrates a typical implementation of data exchange between two MPI processes using \texttt{PMI\_Put}, \texttt{PMI\_Barrier}, and \texttt{PMI\_Get}. In this design, each PMI proxy maintains its own key/value store (KVS), and synchronization is achieved during the collective \texttt{PMI\_Barrier}. Both \texttt{PMI\_Put} and \texttt{PMI\_Get} are local operations that communicate only with the PMI proxy on the same node. \texttt{PMI\_Barrier} is a global collective that requires participation from all MPI processes.
This approach is efficient in scenarios where all processes are collectively exchanging data—only a single \texttt{PMI\_Barrier} is needed to propagate all key/value pairs across the system.

An alternative design may keep a single copy of KVS in the PMI server, requiring each \texttt{PMI\_Put} and \texttt{PMI\_Get} to traverse to the server. While this eliminates the need for a global barrier, it introduces a scalability issue: completing an all-to-all exchange would require $O(P^2)$ point-to-point traversals, where $P$ is the number of processes. Furthermore, a \texttt{PMI\_Get} call may block if the corresponding \texttt{PMI\_Put} has not yet been completed by the sender.

% PMI-2
The PMI-1 interface is relatively simple. In 2010, PMI-2 was proposed to improve scalability. It introduced two key features. First, PMI-2 supports an optional node scope for its key-value store, allowing intra-node \texttt{Put} and \texttt{Get} operations to bypass global synchronization and avoid the use of barriers. This made intra-node data exchange significantly more efficient by avoiding the need for inter-proxy synchronization. Second, PMI-2 allows concurrent PMI operations from multiple threads, potentially improving performance for MPI$+$threads applications.
However, PMI-2 redefined the PMI interface rather than evolving it incrementally from PMI-1, creating adoption challenges. While MPICH supports both PMI-1 and PMI-2, PMI-1 remains the default. Interestingly, the scalability concerns that motivated PMI-2 never fully materialized. Intra-node communication can be efficiently handled via shared memory, bypassing PMI, and multithreaded access to PMI is avoided by restricting its use to the initialization stage, where concurrency is usually not required.
Interestingly, \sessions\ allows the possibility of calling \texttt{MPI\_Session\_init} or the communicator creation routines from concurrent threads, thus requiring multithread support in PMI.
MPICH's support for multithreaded usage of \sessions\ requires an extension to PMI-1 that allows concurrent barrier operations distinguished using string tags. %MW: "is limited" is vague here. Limited how? Is it supported at all? HZ: fixed by stating requirement of PMI extensions. HZ: updated

% PMIx
During the early push toward exascale computing, PMI's scalability limitations again came under scrutiny. It was anticipated that exascale systems might involve 50,000 to 100,000 nodes~\cite{castain2017pmix}, where PMI-1's barrier-based data exchange could significantly delay job startup. In response, PMIx was introduced in 2017~\cite{castain2017pmix}. Although it is presented as an extension of PMI, PMIx has a much broader scope, addressing tasks from efficient executable loading to supporting multiple programming models beyond MPI.
Comparing PMIx directly to PMI-1 or PMI-2 is not straightforward. As noted in its original proposal, PMIx's data exchange interfaces were designed primarily for compatibility with legacy systems. While its scope may have since evolved, it remains unclear whether PMIx fundamentally addresses the core limitations of PMI-1.

Currently, only a subset of the PMIx specification, primarily those features that parallel PMI-1, is required to support MPI implementations.
However, as future versions of MPI introduce capabilities such as fault tolerance and system malleability, broader PMIx functionality may become increasingly important. 
While MPICH will continue to support the PMI-1 interface, extensions may be necessary to acquire certain advanced PMIx features.

% -----------------------------------------
%MW: Similar feedback here: I think this section can be significantly shortened to focus on the background necessary to understand this paper.
\subsection{Efficient MPI Startup}
The efficiency of launching large-scale MPI jobs is influenced by several factors, including the workload manager, resource manager, shared file system, network infrastructure, and the MPI initialization process itself. In this paper, we focus specifically on the MPI layer. Once the MPI processes are instantiated, each process must gather local information and exchange it with its peers—typically via PMI.
A straightforward implementation of this process might look as follows:

\begin{lstlisting}
    PMI_Put(rank, myaddr);
    PMI_Barrier();
    for (int i = 0; i < size; i++) {
        PMI_Get(i, &addrs[i]);
    }
\end{lstlisting}

The PMI interface is designed for simplicity, but its implementations are often not optimized for large-scale environments. For instance, MPICH uses ASCII-based wire protocols, and communication between PMI proxies and the PMI server frequently occurs over TCP connections using the SSH protocol. While this approach is sufficient for medium-scale jobs (e.g., a few thousand processes), it becomes a bottleneck at higher scales. In scenarios with high process-per-node (PPN) counts and thousands of nodes, the address exchange phase can take tens of minutes, significantly impacting total job runtime.

A key insight is that on modern exascale systems, a significant amount of parallelism resides within nodes. In large jobs, the total number of processes is often driven more by high PPN than by the number of nodes. Consequently, intra-node address exchanges can be efficiently handled via shared memory rather than PMI. By limiting PMI communication to one ``node root'' per node and performing the rest of the exchange via shared memory, the overall time spent in \texttt{MPI\_Init} can be drastically reduced to negligible levels~\cite{raffenetti2018locality}.

%MW: I would make this the first subsection of the background.
\subsection{The MPI Sessions API}
The MPI Sessions API, introduced in the MPI-4 standard, provides a more flexible approach to initializing and managing MPI environments. Unlike traditional initialization via \texttt{MPI\_Init} or \texttt{MPI\_Init\_thread}, which implicitly sets up \texttt{MPI\_COMM\_WORLD}, the Sessions model replaces it with the following pseudo code:

\begin{lstlisting}
 MPI_Session_init(info, errhan, &session);
 MPI_Group_from_session_pset(session, "mpi://WORLD", &group);
 MPI_Comm_create_from_group(group, "stringtag", info, errhan, &comm);
\end{lstlisting}

Both \texttt{MPI\_Session\_init} and group preparations are local calls, while \linebreak  \texttt{MPI\_Comm\_create\_from\_group} is collective over the participating process group. In the rest of the paper, we refer to \texttt{MPI\_Session\_init} as local initialization and \texttt{MPI\_Comm\_create\_from\_group} as communicator bootstrapping.

\section{Implementation}
MPICH is a mature project with many downstream vendor implementations, including Cray MPICH~\cite{craympi}, Intel MPI~\cite{intelmpi}, ParaStation MPI~\cite{psmpi}, and MVAPICH~\cite{mvapich}.
When introducing major features such as \sessions, it can be tempting to start with a clean-slate design and backport functionality as needed. This strategy was previously used in the development of shared-memory communication in the CH3 device~\cite{nemesis2007} and lightweight communication in the CH4 device~\cite{guo2025preparing}, helping to minimize disruption to production environments and downstream vendors.
However, starting from scratch often results in long transition periods, during which prior features must be re-implemented, past mistakes may be repeated, and hard-earned lessons must be re-learned. To avoid this, we adopted a strategy of gradual refactoring. Ultimately, we achieved a new design that enabled the implementation of \sessions\ within the existing codebase.
This incremental approach is validated through comprehensive continuous integration (CI) testing and iterative feedback from downstream users and vendors, ensuring both stability and forward progress.

The remainder of this section outlines the key stages involved in preparing for and ultimately implementing full \sessions\ support in MPICH.

% ----
\subsection{Separating Local and Collective Initialization}
The functions \texttt{MPI\_Init} and \texttt{MPI\_Init\_thread} are collective and must be called by all processes, whereas \texttt{MPI\_Session\_init} is a local operation. Therefore, the first step in implementing \sessions\ is to disentangle local-only initializations—suitable for \texttt{MPI\_Session\_init}—from those that require synchronization across processes.
To do this, we split each component’s initialization into two distinct phases: a local phase and a collective phase. While conceptually simple, this separation introduces implementation complexities and can be tedious to apply uniformly across all components.
There is some discretion in deciding which operations belong to the local phase. For instance, initializing self-communication could, in principle, be handled locally, since it does not depend on other processes. However, from a semantic standpoint, it may be more appropriate to delay this step until the creation of the first communicator, as establishing self-communication parallels connecting with a peer. In our implementation, we chose to move as much logic as possible into the local phase to ease the later task of eliminating the dependency on \WORLD.

This separation of local and collective initialization enabled us to implement the complete \sessions\ API in MPICH 4.0. However, because the collective phase still involves constructing a default global context, every process must participate in the first non-self call to \texttt{MPI\_Comm\_create\_from \_group}, during which an internal world communicator is established.
While it supports the majority of applications using \sessions, it retains a hidden dependency on the world communicator.
The remaining subsections describe recent work that eliminates this dependency and realizes true \sessions\ support. This improved implementation is expected to be included in the next MPICH release.

% ----
\subsection{Communicator-Independent Process IDs}

In traditional MPICH implementations, each device defines its own process ID mechanism, typically derived from internal communication data structures. These IDs are used to access communication metadata, such as remote addresses and connection states, resulting in tight coupling between process identity and device internals.
To enable true support for \sessions—without relying on \world—a process ID scheme must be available before device-layer data structures are initialized.

To this end, MPICH introduces a communicator-independent process ID scheme defined at the MPIR layer. Each process is uniquely identified by a pair: a world ID and a world rank.
World Rank corresponds to the process's PMI ID, which, in a traditional world model, is the same as its rank in \WORLD.
World ID is a sequential integer assigned locally by each process. World ID 0 always refers to the process's original world, where the process is created in. New worlds encountered via dynamic process creation or internal mechanisms are assigned incrementally higher world IDs.
A world is defined as a group of processes sharing a PMI namespace---a logical unit of coordination that shares a key-value store. In the world model, a world consists of processes that make up \WORLD.

A key step in constructing communicators is exchanging process metadata, either via PMI or through a direct connection established via the dynamic process interface. When all process IDs originate from the same PMI interface, they can be directly mapped to PMI identifiers (e.g., PMI namespace and PMI IDs) for metadata exchange.
However, dynamic processes may not share a common PMI context. Therefore, MPICH must explicitly exchange world information during dynamic process operations (e.g., \texttt{MPI\_Accept} and \texttt{MPI\_Connect}). This ensures proper mapping between remote and local world IDs and allows processes to populate their local world tables with the appropriate metadata.
This design enables MPICH to support a wide range of PMI implementations and dynamic process scenarios, providing flexibility while maintaining consistency.

MPICH's device layers have been modified to adopt this MPIR-defined process ID scheme. While this change reduces the device layer’s flexibility in designing its own communication databases, it eliminates redundant implementations and promotes consistency across devices by offloading ID management to a centralized, device-independent layer.

% ----
\subsection{Group-Collective Address Exchange Over PMI}

A key step in bootstrapping communicators is the exchange of network addresses using PMI. In the traditional \WORLD\ model, all processes collectively exchange addresses, as described in Section~\ref{PMI}. However, in the \sessions\ model, a global initialization stage may not exist. Instead, users may choose to bootstrap a communicator over a subset of processes rather than the entire process set.

MPICH can be configured to work with different PMI interfaces including PMI-1, PMI-2, and PMIx. 
Group-level data exchange in PMIx is supported via \texttt{PMIx\_Put}, \texttt{PMIx\_Get}, and \texttt{PMIx\_Fence}. In particular, \texttt{PMIx\_Fence} accepts a process array as an argument, allowing it to execute a blocking barrier across a specified group while collecting data posted via \texttt{PMIx\_Put}.
However, support for performing \texttt{PMIx\_Fence} over subsets of \WORLD\ remains inconsistent across implementations.
In testing our implementation on the Cray Programming Environment using the Parallel Application Launch Service (PALS), we observed that jobs could hang or crash.
As a workaround until the PMIx implementation issues are fixed, we introduce a fallback environment variable, which when set, all processes must call \texttt{MPI\_Comm\_create\_from\_group} collectively, allowing \texttt{PMIx\_Fence} to be executed over \WORLD. Processes not part of the target group participate only in the \texttt{PMIx\_Fence} call to ensure progress and consistency.

To support group-level data exchange in PMI-1 or PMI-2, extensions are required. We propose a new collective operation, \texttt{PMI\_Barrier\_group}:
\begin{lstlisting}
    int PMI_Barrier_group(const int *group, int count);
\end{lstlisting}
This extension achieves the same semantics as \texttt{PMIx\_Fence} over a group of processes.
When using a PMI server that supports this extension, such as MPICH's bundled process launcher, Hydra, there won't be restrictions. Otherwise, the same fallback environment variable must be set and all processes are required to call \texttt{MPI\_Comm\_create\_from\_group} collectively.

An additional PMI extension is required to support multithreaded usage. However, since most applications do not require concurrent access to communicator bootstrapping, we defer this work to future development.

% ----
\subsection{Atomic Shared Memory Initialization}
Unlike network communications, where remote connections can be established on demand, shared memory communication requires all participating processes to share a common memory segment. The segment must be large enough to accommodate all local processes on the node even if some are not  present during the initial communicator bootstrapping phase.
Additionally, each process must ensure that its peers are ready before using the shared memory for communication.
In previous versions of MPICH, which includes an explicit global initialization stage, shared memory initialization is accomplished via collectives in which all local processes participate. 
To support the true Sessions model, however, we cannot rely on collectives since not all local processes may be present during the communicator bootstrapping. Therefore, we redesigned shared memory initialization based on atomic operations.

Atomic shared memory creation is performed using the POSIX API \texttt{shm\_open}. Each process initially calls \texttt{shm\_open} with the flag \texttt{O\_CREAT | O\_EXCL}. The first process to succeed in this call becomes the ``root'' and is responsible for creating and initializing the shared memory segment. All other processes, upon failing the initial call, subsequently invoke \texttt{shm\_open} again—this time to open the segment created by the root.
Other processes must wait for the root to complete initializing the memory. This is coordinated in two steps. First, processes call \texttt{fstat} on the file descriptor returned by \texttt{shm\_open} to ensure that the root process has set the size of the region before mapping it. Second, processes read an atomic variable named \texttt{root\_ready} from a fixed location in the shared memory region until the value indicates the root process has completed setup.
In addition, each process maintains its own atomic \texttt{ready} flag. During a communicator bootstrapping call, each process waits for the \texttt{ready} flags of all other local processes in the group before completing the call, ensuring the communicator is fully initialized and safe to use upon return.

% ----
\subsection{Bootstrapping Communicators from Sparse Connections}
PMI is designed primarily for bootstrapping MPI and emphasizes simplicity over communication efficiency. While this approach is adequate for small to medium-sized jobs involving thousands of processes, it becomes unsuitable for large-scale all-to-all address exchanges. In jobs with hundreds of thousands of processes—often with high process-per-node (PPN) counts—using PMI for such exchanges can make MPI initialization prohibitively slow.
To improve scalability and efficiency, we use PMI only to exchange addresses between node roots within the target communicator. Once this exchange is complete and connections are established, a sub-communicator consisting of the node roots—referred to as \texttt{node\_roots\_comm}—becomes operational.
Within each node, after local processes initialize shared memory, another sub-communicator composed of all local processes—referred to as \texttt{node\_comm}—is also brought online. With both \texttt{node\_roots\_comm} and \texttt{node\_comm} active, communication paths are established between any pair of processes. At this point, the communicator forms a sparsely connected graph rather than a fully connected (logical all-to-all) topology.
In the final phase, we perform an \texttt{MPI\_Allgather} across all processes to exchange address information and establish a logical all-to-all connection. This step is executed using MPI collectives instead of PMI, as MPI can take advantage of high-bandwidth interconnects such as InfiniBand and shared memory, providing significantly better performance.
Because the communicator begins in a sparsely connected state, we employ a specialized hierarchical collective algorithm to efficiently perform the \texttt{MPI\_Allgather}. After this stage, the communicator is fully bootstrapped and ready for use.

While previous versions of MPICH included a similar algorithm that used node roots to bootstrap all-to-all address exchange, it assumed a world initialization context. In this work, we redesigned the algorithm to support arbitrary communicator bootstrapping.
\section{Experimental Evaluation}

MPICH’s implementation of \sessions\ modifies only the MPI initialization process and the bootstrapping of initial communicators. Runtime communication behavior remains unchanged between the world and Session models. Therefore, our evaluation focuses on the scalability of initialization, specifically comparing the world model against the Session model.
All experiments were conducted on the Aurora system at the Argonne Leadership Computing Facility~\cite{aurora}, using between 1 and 2048 nodes, with 96 processes per node (PPN). Each node is equipped with two Intel Xeon CPU Max Series processors, 512~GB of DDR5 RAM, six Intel Data Center GPUs, and eight Slingshot 11 network endpoints.
To ensure a clear measurement of MPI initialization scalability, MPICH’s GPU support was disabled, as GPU initialization introduces significant time and memory overhead that can obscure the interesting part of the measurements.
A high PPN configuration was chosen to emphasize scalability challenges. Although additional experiments were performed with 12 PPN, those results are omitted due to space constraints.

\subsection{Comparing World Initialization}
Traditional MPI applications can be adapted to use \sessions\ by replacing calls to \texttt{MPI\_Init} or \texttt{MPI\_Init\_thread} with an initial call to \texttt{MPI\_Session\_init}, followed by \texttt{MPI\_Comm\_create\_from\_group} using groups derived from the process sets ``mpi://SELF'' and ``mpi://WORLD''. These steps create communicators functionally equivalent to \texttt{MPI\_COMM\_SELF} and \WORLD, respectively.
In MPICH, communicators created via \sessions\ behave identically to those created under the traditional world model.

% The session measurements are divided into 3 phases: MPI\_Session\_init, creating communicator from ``mpi://SELF'', and creating communicator from ``mpi://WORLD''.

\begin{figure} 
    \centering
\resizebox{0.95\textwidth}{!}{
    \begin{tikzpicture}[
    dummy/.style={fill=blue!80!gray},
    sess_init/.style={fill=orange},
    sess_self/.style={fill=purple},
    sess_world/.style={fill=green},
]
% title
\node at (5, 6.3) {\large (a) MPICH Dev - Init Time};
% x labels
\draw (0.4167, 0-0.5) node {1};
\draw (1.25, 0-0.5) node {2};
\draw (2.083, 0-0.5) node {4};
\draw (2.917, 0-0.5) node {8};
\draw (3.75, 0-0.5) node {16};
\draw (4.583, 0-0.5) node {32};
\draw (5.417, 0-0.5) node {64};
\draw (6.25, 0-0.5) node {128};
\draw (7.083, 0-0.5) node {256};
\draw (7.917, 0-0.5) node {512};
\draw (8.75, 0-0.5) node {1024};
\draw (9.583, 0-0.5) node {2048};
% y labels
\draw (0-0.3, 0) node[anchor=east] {0};
\draw[gray] (0, 0) -- (10, 0);
\draw (0-0.3, 0.7333) node[anchor=east] {2};
\draw[gray] (0, 0.7333) -- (10, 0.7333);
\draw (0-0.3, 1.467) node[anchor=east] {4};
\draw[gray] (0, 1.467) -- (10, 1.467);
\draw (0-0.3, 2.2) node[anchor=east] {6};
\draw[gray] (0, 2.2) -- (10, 2.2);
\draw (0-0.3, 2.933) node[anchor=east] {8};
\draw[gray] (0, 2.933) -- (10, 2.933);
\draw (0-0.3, 3.667) node[anchor=east] {10};
\draw[gray] (0, 3.667) -- (10, 3.667);
\draw (0-0.3, 4.4) node[anchor=east] {12};
\draw[gray] (0, 4.4) -- (10, 4.4);
\draw (0-0.3, 5.133) node[anchor=east] {14};
\draw[gray] (0, 5.133) -- (10, 5.133);
\node at (-0.5, 5.6) {sec};
\node at (5, -1) {Number of nodes};
% legend
\node[dummy, minimum width=0.3, minimum height=0.3] at (0.5, -2) {};
\node[anchor=west] at (0.5+0.2, -2) {MPI\_Init};
\node[sess_init, minimum width=0.3, minimum height=0.3] at (2.7, -2) {};
\node[anchor=west] at (2.7+0.2, -2) {Session Init};
\node[sess_self, minimum width=0.3, minimum height=0.3] at (5.3, -2) {};
\node[anchor=west] at (5.3+0.2, -2) {Self Comm};
\node[sess_world, minimum width=0.3, minimum height=0.3] at (7.6, -2) {};
\node[anchor=west] at (7.6+0.2, -2) {World Comm};
% MPI_Init
\fill[dummy] (0.167, 0) rectangle (0.383, 1.5);
\fill[sess_init] (0.45, 0) rectangle (0.667, 1.47);
\fill[sess_self] (0.45, 1.47) rectangle (0.667, 1.48);
\fill[sess_world] (0.45, 1.48) rectangle (0.667, 1.5);
\fill[dummy] (1, 0) rectangle (1.22, 1.52);
\fill[sess_init] (1.28, 0) rectangle (1.5, 1.5);
\fill[sess_self] (1.28, 1.5) rectangle (1.5, 1.51);
\fill[sess_world] (1.28, 1.51) rectangle (1.5, 1.52);
\fill[dummy] (1.83, 0) rectangle (2.05, 1.58);
\fill[sess_init] (2.12, 0) rectangle (2.33, 1.54);
\fill[sess_self] (2.12, 1.54) rectangle (2.33, 1.55);
\fill[sess_world] (2.12, 1.55) rectangle (2.33, 1.56);
\fill[dummy] (2.67, 0) rectangle (2.88, 1.54);
\fill[sess_init] (2.95, 0) rectangle (3.17, 1.51);
\fill[sess_self] (2.95, 1.51) rectangle (3.17, 1.52);
\fill[sess_world] (2.95, 1.52) rectangle (3.17, 1.53);
\fill[dummy] (3.5, 0) rectangle (3.72, 1.55);
\fill[sess_init] (3.78, 0) rectangle (4, 1.51);
\fill[sess_self] (3.78, 1.51) rectangle (4, 1.52);
\fill[sess_world] (3.78, 1.52) rectangle (4, 1.54);
\fill[dummy] (4.33, 0) rectangle (4.55, 1.6);
\fill[sess_init] (4.62, 0) rectangle (4.83, 1.58);
\fill[sess_self] (4.62, 1.58) rectangle (4.83, 1.59);
\fill[sess_world] (4.62, 1.59) rectangle (4.83, 1.62);
\fill[dummy] (5.17, 0) rectangle (5.38, 1.66);
\fill[sess_init] (5.45, 0) rectangle (5.67, 1.61);
\fill[sess_self] (5.45, 1.61) rectangle (5.67, 1.62);
\fill[sess_world] (5.45, 1.62) rectangle (5.67, 1.69);
\fill[dummy] (6, 0) rectangle (6.22, 1.76);
\fill[sess_init] (6.28, 0) rectangle (6.5, 1.66);
\fill[sess_self] (6.28, 1.66) rectangle (6.5, 1.67);
\fill[sess_world] (6.28, 1.67) rectangle (6.5, 1.83);
\fill[dummy] (6.83, 0) rectangle (7.05, 1.88);
\fill[sess_init] (7.12, 0) rectangle (7.33, 1.7);
\fill[sess_self] (7.12, 1.7) rectangle (7.33, 1.71);
\fill[sess_world] (7.12, 1.71) rectangle (7.33, 2.02);
\fill[dummy] (7.67, 0) rectangle (7.88, 2.13);
\fill[sess_init] (7.95, 0) rectangle (8.17, 1.79);
\fill[sess_self] (7.95, 1.79) rectangle (8.17, 1.8);
\fill[sess_world] (7.95, 1.8) rectangle (8.17, 2.43);
\fill[dummy] (8.5, 0) rectangle (8.72, 2.56);
\fill[sess_init] (8.78, 0) rectangle (9, 1.95);
\fill[sess_self] (8.78, 1.95) rectangle (9, 1.97);
\fill[sess_world] (8.78, 1.97) rectangle (9, 3.42);
\fill[dummy] (9.33, 0) rectangle (9.55, 3.87);
\fill[sess_init] (9.62, 0) rectangle (9.83, 2.42);
\fill[sess_self] (9.62, 2.42) rectangle (9.83, 2.42);
\fill[sess_world] (9.62, 2.42) rectangle (9.83, 5.35);
% title
\node at (17, 6.3) {\large (b) MPICH Dev - Node Memory};
% x labels
\draw (12.42, 0-0.5) node {1};
\draw (13.25, 0-0.5) node {2};
\draw (14.08, 0-0.5) node {4};
\draw (14.92, 0-0.5) node {8};
\draw (15.75, 0-0.5) node {16};
\draw (16.58, 0-0.5) node {32};
\draw (17.42, 0-0.5) node {64};
\draw (18.25, 0-0.5) node {128};
\draw (19.08, 0-0.5) node {256};
\draw (19.92, 0-0.5) node {512};
\draw (20.75, 0-0.5) node {1024};
\draw (21.58, 0-0.5) node {2048};
% y labels
\draw (12-0.3, 0) node[anchor=east] {0};
\draw[gray] (12, 0) -- (22, 0);
\draw (12-0.3, 0.6111) node[anchor=east] {2};
\draw[gray] (12, 0.6111) -- (22, 0.6111);
\draw (12-0.3, 1.222) node[anchor=east] {4};
\draw[gray] (12, 1.222) -- (22, 1.222);
\draw (12-0.3, 1.833) node[anchor=east] {6};
\draw[gray] (12, 1.833) -- (22, 1.833);
\draw (12-0.3, 2.444) node[anchor=east] {8};
\draw[gray] (12, 2.444) -- (22, 2.444);
\draw (12-0.3, 3.056) node[anchor=east] {10};
\draw[gray] (12, 3.056) -- (22, 3.056);
\draw (12-0.3, 3.667) node[anchor=east] {12};
\draw[gray] (12, 3.667) -- (22, 3.667);
\draw (12-0.3, 4.278) node[anchor=east] {14};
\draw[gray] (12, 4.278) -- (22, 4.278);
\draw (12-0.3, 4.889) node[anchor=east] {16};
\draw[gray] (12, 4.889) -- (22, 4.889);
\draw (12-0.3, 5.5) node[anchor=east] {18};
\draw[gray] (12, 5.5) -- (22, 5.5);
\node at (11.5, 6) {GB};
\node at (17, -1) {Number of nodes};
% legend
\node[dummy, minimum width=0.3, minimum height=0.3] at (12.5, -2) {};
\node[anchor=west] at (12.5+0.2, -2) {MPI\_Init};
\node[sess_init, minimum width=0.3, minimum height=0.3] at (14.7, -2) {};
\node[anchor=west] at (14.7+0.2, -2) {Session Init};
\node[sess_self, minimum width=0.3, minimum height=0.3] at (17.3, -2) {};
\node[anchor=west] at (17.3+0.2, -2) {Self Comm};
\node[sess_world, minimum width=0.3, minimum height=0.3] at (19.6, -2) {};
\node[anchor=west] at (19.6+0.2, -2) {World Comm};
% MPI_Init
\fill[dummy] (12.2, 0) rectangle (12.4, 2.27);
\fill[sess_init] (12.5, 0) rectangle (12.7, 2.19);
\fill[sess_self] (12.5, 2.19) rectangle (12.7, 2.25);
\fill[sess_world] (12.5, 2.25) rectangle (12.7, 2.26);
\fill[dummy] (13, 0) rectangle (13.2, 2.26);
\fill[sess_init] (13.3, 0) rectangle (13.5, 2.19);
\fill[sess_self] (13.3, 2.19) rectangle (13.5, 2.25);
\fill[sess_world] (13.3, 2.25) rectangle (13.5, 2.26);
\fill[dummy] (13.8, 0) rectangle (14, 2.26);
\fill[sess_init] (14.1, 0) rectangle (14.3, 2.19);
\fill[sess_self] (14.1, 2.19) rectangle (14.3, 2.24);
\fill[sess_world] (14.1, 2.24) rectangle (14.3, 2.25);
\fill[dummy] (14.7, 0) rectangle (14.9, 2.29);
\fill[sess_init] (15, 0) rectangle (15.2, 2.2);
\fill[sess_self] (15, 2.2) rectangle (15.2, 2.26);
\fill[sess_world] (15, 2.26) rectangle (15.2, 2.27);
\fill[dummy] (15.5, 0) rectangle (15.7, 2.29);
\fill[sess_init] (15.8, 0) rectangle (16, 2.21);
\fill[sess_self] (15.8, 2.21) rectangle (16, 2.27);
\fill[sess_world] (15.8, 2.27) rectangle (16, 2.28);
\fill[dummy] (16.3, 0) rectangle (16.5, 2.3);
\fill[sess_init] (16.6, 0) rectangle (16.8, 2.21);
\fill[sess_self] (16.6, 2.21) rectangle (16.8, 2.27);
\fill[sess_world] (16.6, 2.27) rectangle (16.8, 2.28);
\fill[dummy] (17.2, 0) rectangle (17.4, 2.38);
\fill[sess_init] (17.5, 0) rectangle (17.7, 2.27);
\fill[sess_self] (17.5, 2.27) rectangle (17.7, 2.33);
\fill[sess_world] (17.5, 2.33) rectangle (17.7, 2.35);
\fill[dummy] (18, 0) rectangle (18.2, 2.45);
\fill[sess_init] (18.3, 0) rectangle (18.5, 2.37);
\fill[sess_self] (18.3, 2.37) rectangle (18.5, 2.42);
\fill[sess_world] (18.3, 2.42) rectangle (18.5, 2.44);
\fill[dummy] (18.8, 0) rectangle (19, 2.63);
\fill[sess_init] (19.1, 0) rectangle (19.3, 2.54);
\fill[sess_self] (19.1, 2.54) rectangle (19.3, 2.59);
\fill[sess_world] (19.1, 2.59) rectangle (19.3, 2.63);
\fill[dummy] (19.7, 0) rectangle (19.9, 3.03);
\fill[sess_init] (20, 0) rectangle (20.2, 2.9);
\fill[sess_self] (20, 2.9) rectangle (20.2, 3.01);
\fill[sess_world] (20, 3.01) rectangle (20.2, 3.08);
\fill[dummy] (20.5, 0) rectangle (20.7, 3.81);
\fill[sess_init] (20.8, 0) rectangle (21, 3.6);
\fill[sess_self] (20.8, 3.6) rectangle (21, 3.69);
\fill[sess_world] (20.8, 3.69) rectangle (21, 3.81);
\fill[dummy] (21.3, 0) rectangle (21.5, 5.37);
\fill[sess_init] (21.6, 0) rectangle (21.8, 5.01);
\fill[sess_self] (21.6, 5.01) rectangle (21.8, 5.08);
\fill[sess_world] (21.6, 5.08) rectangle (21.8, 5.35);
\end{tikzpicture}
}    
\caption{Comparing MPI Initialization between the world model and the session model using mpich-dev. The session model measurements are split into session init and bootstrapping the self and the world communicators. (a) Initialization times in seconds. (b) Node memory usage in GB. }
\label{mpich-dev}
\end{figure}
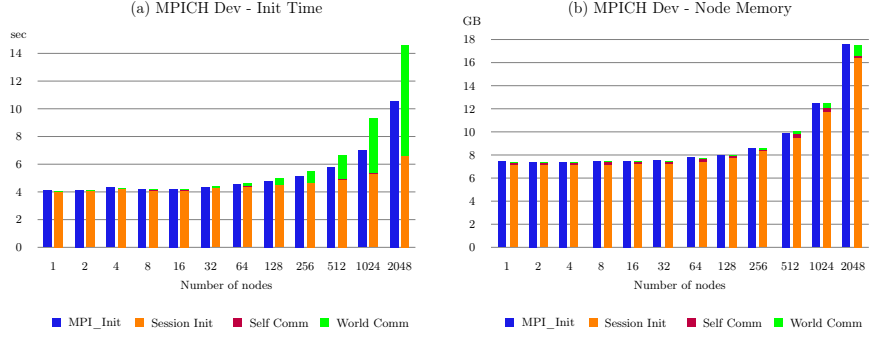

Figure~\ref{mpich-dev} presents the initialization time and memory usage for the development version of MPICH (MPICH-dev).
Initialization time is calculated as the average, across all processes, of the duration taken to execute \texttt{MPI\_Init}, \texttt{MPI\_Session\_init}, and \texttt{MPI\_Comm\_create\_from\_group}, respectively.
Memory usage is estimated based on the \texttt{MemFree} field from \texttt{/proc/meminfo}, capturing the total memory consumed by all processes on a node.
To minimize interference from phase misalignment among processes on the same node, deliberate pauses are inserted between measurement phases.

The results reveal several key observations. Local initialization, represented by \SessionInit, accounts for the majority of both initialization time and memory usage. Only a small fraction of this cost is due to MPICH itself. The bulk originates from lower-level components such as libfabric and hardware supporting libraries. The overhead is likely due to slow hardware initialization on hybrid-node architectures with multiple GPUs and NICs, with performance further degraded under high PPN as processes compete for resources. Additional overhead may come from dynamic loading of supporting libraries, which stresses shared file systems at scale.
The memory allocations may include buffers anticipating a full world communication, which may explain why the local initialization also increases in both the initialization time and memory consumption as the total number of nodes increase.

As an anecdote, MPICH defaults to using the \texttt{tcp;rxm} libfabric provider instead of the \texttt{cxi} provider when a job runs on a single node. This choice avoids the resource limitations associated with the \texttt{cxi} provider for single-node jobs.
Our initial experiments showed a memory usage outlier at the 1-node scale, caused by \texttt{tcp;rxm} allocating significantly more memory (~17GB) than \texttt{cxi}. By explicitly selecting the \texttt{cxi} provider, we obtained the consistent measurements shown in Fig.~\ref{mpich-dev}.

Creating the self communicator incurs a negligible, constant cost. In contrast, constructing the world communicator, the only collective step during initialization, scales with node count in both time and memory, though it only becomes a significant factor at scales above 1024 nodes.

Overall, session-based initialization takes slightly longer than the world model, while both models exhibit comparable memory usage.
The reason for the longer initialization time in the sessions model is not entirely clear. One contributing factor may be the additional context ID allocation required during communicator bootstrapping in the sessions model.
However, we suspect the more significant factor is overhead caused by process imbalance. In the world model, local initialization and world communicator construction are combined, which may help mask some of the load imbalance.

\subsection{Comparing Implementations}
To provide a broader comparison, we conducted the same set of experiments using MPICH 4.3.0 and Open MPI 5.0.7.

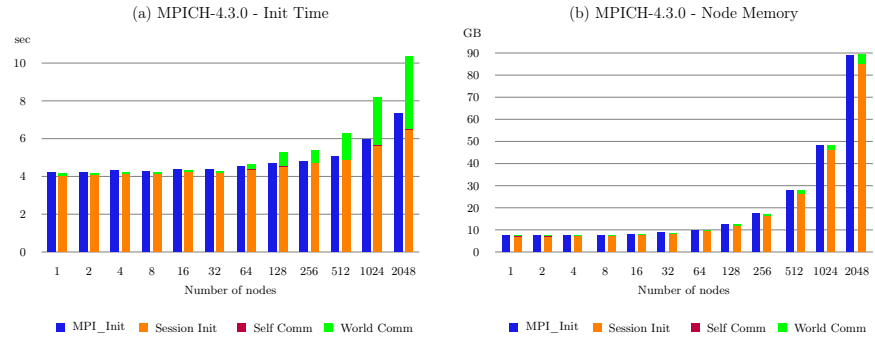
\begin{figure}
    \centering
\resizebox{0.95\textwidth}{!}{
    \begin{tikzpicture}[
    dummy/.style={fill=blue!80!gray},
    sess_init/.style={fill=orange},
    sess_self/.style={fill=purple},
    sess_world/.style={fill=green},
]
% title
\node at (5, 6.3) {\large (a) MPICH-4.3.0 - Init Time};
% x labels
\draw (0.4167, 0-0.5) node {1};
\draw (1.25, 0-0.5) node {2};
\draw (2.083, 0-0.5) node {4};
\draw (2.917, 0-0.5) node {8};
\draw (3.75, 0-0.5) node {16};
\draw (4.583, 0-0.5) node {32};
\draw (5.417, 0-0.5) node {64};
\draw (6.25, 0-0.5) node {128};
\draw (7.083, 0-0.5) node {256};
\draw (7.917, 0-0.5) node {512};
\draw (8.75, 0-0.5) node {1024};
\draw (9.583, 0-0.5) node {2048};
% y labels
\draw (0-0.3, 0) node[anchor=east] {0};
\draw[gray] (0, 0) -- (10, 0);
\draw (0-0.3, 1) node[anchor=east] {2};
\draw[gray] (0, 1) -- (10, 1);
\draw (0-0.3, 2) node[anchor=east] {4};
\draw[gray] (0, 2) -- (10, 2);
\draw (0-0.3, 3) node[anchor=east] {6};
\draw[gray] (0, 3) -- (10, 3);
\draw (0-0.3, 4) node[anchor=east] {8};
\draw[gray] (0, 4) -- (10, 4);
\draw (0-0.3, 5) node[anchor=east] {10};
\draw[gray] (0, 5) -- (10, 5);
% legend
\node at (-0.5, 5.6) {sec};
\node at (5, -1) {Number of nodes};
\node[dummy, minimum width=0.3, minimum height=0.3] at (0.5, -2) {};
\node[anchor=west] at (0.5+0.2, -2) {MPI\_Init};
\node[sess_init, minimum width=0.3, minimum height=0.3] at (2.7, -2) {};
\node[anchor=west] at (2.7+0.2, -2) {Session Init};
\node[sess_self, minimum width=0.3, minimum height=0.3] at (5.3, -2) {};
\node[anchor=west] at (5.3+0.2, -2) {Self Comm};
\node[sess_world, minimum width=0.3, minimum height=0.3] at (7.6, -2) {};
\node[anchor=west] at (7.6+0.2, -2) {World Comm};
% MPI_Init
\fill[dummy] (0.167, 0) rectangle (0.383, 2.1);
\fill[sess_init] (0.45, 0) rectangle (0.667, 2.02);
\fill[sess_self] (0.45, 2.02) rectangle (0.667, 2.02);
\fill[sess_world] (0.45, 2.02) rectangle (0.667, 2.08);
\fill[dummy] (1, 0) rectangle (1.22, 2.11);
\fill[sess_init] (1.28, 0) rectangle (1.5, 2.04);
\fill[sess_self] (1.28, 2.04) rectangle (1.5, 2.04);
\fill[sess_world] (1.28, 2.04) rectangle (1.5, 2.08);
\fill[dummy] (1.83, 0) rectangle (2.05, 2.15);
\fill[sess_init] (2.12, 0) rectangle (2.33, 2.08);
\fill[sess_self] (2.12, 2.08) rectangle (2.33, 2.08);
\fill[sess_world] (2.12, 2.08) rectangle (2.33, 2.12);
\fill[dummy] (2.67, 0) rectangle (2.88, 2.13);
\fill[sess_init] (2.95, 0) rectangle (3.17, 2.07);
\fill[sess_self] (2.95, 2.07) rectangle (3.17, 2.07);
\fill[sess_world] (2.95, 2.07) rectangle (3.17, 2.11);
\fill[dummy] (3.5, 0) rectangle (3.72, 2.19);
\fill[sess_init] (3.78, 0) rectangle (4, 2.12);
\fill[sess_self] (3.78, 2.12) rectangle (4, 2.12);
\fill[sess_world] (3.78, 2.12) rectangle (4, 2.17);
\fill[dummy] (4.33, 0) rectangle (4.55, 2.18);
\fill[sess_init] (4.62, 0) rectangle (4.83, 2.1);
\fill[sess_self] (4.62, 2.1) rectangle (4.83, 2.1);
\fill[sess_world] (4.62, 2.1) rectangle (4.83, 2.15);
\fill[dummy] (5.17, 0) rectangle (5.38, 2.27);
\fill[sess_init] (5.45, 0) rectangle (5.67, 2.19);
\fill[sess_self] (5.45, 2.19) rectangle (5.67, 2.19);
\fill[sess_world] (5.45, 2.19) rectangle (5.67, 2.31);
\fill[dummy] (6, 0) rectangle (6.22, 2.34);
\fill[sess_init] (6.28, 0) rectangle (6.5, 2.27);
\fill[sess_self] (6.28, 2.27) rectangle (6.5, 2.27);
\fill[sess_world] (6.28, 2.27) rectangle (6.5, 2.64);
\fill[dummy] (6.83, 0) rectangle (7.05, 2.4);
\fill[sess_init] (7.12, 0) rectangle (7.33, 2.38);
\fill[sess_self] (7.12, 2.38) rectangle (7.33, 2.38);
\fill[sess_world] (7.12, 2.38) rectangle (7.33, 2.7);
\fill[dummy] (7.67, 0) rectangle (7.88, 2.54);
\fill[sess_init] (7.95, 0) rectangle (8.17, 2.44);
\fill[sess_self] (7.95, 2.44) rectangle (8.17, 2.44);
\fill[sess_world] (7.95, 2.44) rectangle (8.17, 3.15);
\fill[dummy] (8.5, 0) rectangle (8.72, 2.97);
\fill[sess_init] (8.78, 0) rectangle (9, 2.82);
\fill[sess_self] (8.78, 2.82) rectangle (9, 2.82);
\fill[sess_world] (8.78, 2.82) rectangle (9, 4.09);
\fill[dummy] (9.33, 0) rectangle (9.55, 3.67);
\fill[sess_init] (9.62, 0) rectangle (9.83, 3.25);
\fill[sess_self] (9.62, 3.25) rectangle (9.83, 3.25);
\fill[sess_world] (9.62, 3.25) rectangle (9.83, 5.18);
% title
\node at (17, 6.3) {\large (b) MPICH-4.3.0 - Node Memory};
% x labels
\draw (12.42, 0-0.5) node {1};
\draw (13.25, 0-0.5) node {2};
\draw (14.08, 0-0.5) node {4};
\draw (14.92, 0-0.5) node {8};
\draw (15.75, 0-0.5) node {16};
\draw (16.58, 0-0.5) node {32};
\draw (17.42, 0-0.5) node {64};
\draw (18.25, 0-0.5) node {128};
\draw (19.08, 0-0.5) node {256};
\draw (19.92, 0-0.5) node {512};
\draw (20.75, 0-0.5) node {1024};
\draw (21.58, 0-0.5) node {2048};
% y labels
\draw (12-0.3, 0) node[anchor=east] {0};
\draw[gray] (12, 0) -- (22, 0);
\draw (12-0.3, 0.5851) node[anchor=east] {10};
\draw[gray] (12, 0.5851) -- (22, 0.5851);
\draw (12-0.3, 1.17) node[anchor=east] {20};
\draw[gray] (12, 1.17) -- (22, 1.17);
\draw (12-0.3, 1.755) node[anchor=east] {30};
\draw[gray] (12, 1.755) -- (22, 1.755);
\draw (12-0.3, 2.34) node[anchor=east] {40};
\draw[gray] (12, 2.34) -- (22, 2.34);
\draw (12-0.3, 2.926) node[anchor=east] {50};
\draw[gray] (12, 2.926) -- (22, 2.926);
\draw (12-0.3, 3.511) node[anchor=east] {60};
\draw[gray] (12, 3.511) -- (22, 3.511);
\draw (12-0.3, 4.096) node[anchor=east] {70};
\draw[gray] (12, 4.096) -- (22, 4.096);
\draw (12-0.3, 4.681) node[anchor=east] {80};
\draw[gray] (12, 4.681) -- (22, 4.681);
\draw (12-0.3, 5.266) node[anchor=east] {90};
\draw[gray] (12, 5.266) -- (22, 5.266);
% legend
\node at (11.4, 5.8) {GB};
\node at (17, -1) {Number of nodes};
\node[dummy, minimum width=0.3, minimum height=0.3] at (12.5, -2) {};
\node[anchor=west] at (12.5+0.2, -2) {MPI\_Init};
\node[sess_init, minimum width=0.3, minimum height=0.3] at (14.7, -2) {};
\node[anchor=west] at (14.7+0.2, -2) {Session Init};
\node[sess_self, minimum width=0.3, minimum height=0.3] at (17.3, -2) {};
\node[anchor=west] at (17.3+0.2, -2) {Self Comm};
\node[sess_world, minimum width=0.3, minimum height=0.3] at (19.6, -2) {};
\node[anchor=west] at (19.6+0.2, -2) {World Comm};
% MPI_Init
\fill[dummy] (12.2, 0) rectangle (12.4, 0.435);
\fill[sess_init] (12.5, 0) rectangle (12.7, 0.42);
\fill[sess_self] (12.5, 0.42) rectangle (12.7, 0.42);
\fill[sess_world] (12.5, 0.42) rectangle (12.7, 0.434);
\fill[dummy] (13, 0) rectangle (13.2, 0.438);
\fill[sess_init] (13.3, 0) rectangle (13.5, 0.422);
\fill[sess_self] (13.3, 0.422) rectangle (13.5, 0.422);
\fill[sess_world] (13.3, 0.422) rectangle (13.5, 0.436);
\fill[dummy] (13.8, 0) rectangle (14, 0.441);
\fill[sess_init] (14.1, 0) rectangle (14.3, 0.427);
\fill[sess_self] (14.1, 0.427) rectangle (14.3, 0.427);
\fill[sess_world] (14.1, 0.427) rectangle (14.3, 0.441);
\fill[dummy] (14.7, 0) rectangle (14.9, 0.449);
\fill[sess_init] (15, 0) rectangle (15.2, 0.435);
\fill[sess_self] (15, 0.435) rectangle (15.2, 0.435);
\fill[sess_world] (15, 0.435) rectangle (15.2, 0.449);
\fill[dummy] (15.5, 0) rectangle (15.7, 0.47);
\fill[sess_init] (15.8, 0) rectangle (16, 0.454);
\fill[sess_self] (15.8, 0.454) rectangle (16, 0.454);
\fill[sess_world] (15.8, 0.454) rectangle (16, 0.469);
\fill[dummy] (16.3, 0) rectangle (16.5, 0.505);
\fill[sess_init] (16.6, 0) rectangle (16.8, 0.489);
\fill[sess_self] (16.6, 0.489) rectangle (16.8, 0.489);
\fill[sess_world] (16.6, 0.489) rectangle (16.8, 0.506);
\fill[dummy] (17.2, 0) rectangle (17.4, 0.579);
\fill[sess_init] (17.5, 0) rectangle (17.7, 0.56);
\fill[sess_self] (17.5, 0.56) rectangle (17.7, 0.56);
\fill[sess_world] (17.5, 0.56) rectangle (17.7, 0.58);
\fill[dummy] (18, 0) rectangle (18.2, 0.721);
\fill[sess_init] (18.3, 0) rectangle (18.5, 0.704);
\fill[sess_self] (18.3, 0.704) rectangle (18.5, 0.704);
\fill[sess_world] (18.3, 0.704) rectangle (18.5, 0.731);
\fill[dummy] (18.8, 0) rectangle (19, 1.02);
\fill[sess_init] (19.1, 0) rectangle (19.3, 0.971);
\fill[sess_self] (19.1, 0.971) rectangle (19.3, 0.971);
\fill[sess_world] (19.1, 0.971) rectangle (19.3, 1.01);
\fill[dummy] (19.7, 0) rectangle (19.9, 1.62);
\fill[sess_init] (20, 0) rectangle (20.2, 1.56);
\fill[sess_self] (20, 1.56) rectangle (20.2, 1.56);
\fill[sess_world] (20, 1.56) rectangle (20.2, 1.64);
\fill[dummy] (20.5, 0) rectangle (20.7, 2.81);
\fill[sess_init] (20.8, 0) rectangle (21, 2.7);
\fill[sess_self] (20.8, 2.7) rectangle (21, 2.7);
\fill[sess_world] (20.8, 2.7) rectangle (21, 2.83);
\fill[dummy] (21.3, 0) rectangle (21.5, 5.21);
\fill[sess_init] (21.6, 0) rectangle (21.8, 4.99);
\fill[sess_self] (21.6, 4.99) rectangle (21.8, 4.99);
\fill[sess_world] (21.6, 4.99) rectangle (21.8, 5.25);
\end{tikzpicture}
}    
\caption{Comparing MPI Initialization between the world model and the session model using MPICH 4.3.0. (a) Initialization times in seconds. (b) Node memory usage in GB. }
 \label{mpich43}
\end{figure}

Figure~\ref{mpich43} shows the results for MPICH 4.3.0, the current stable release, which implements \sessions\ using an internal world communicator. As a result, it only supports bootstrapping either the self or world communicator. The comparison between the world and session models in MPICH 4.3.0 is similar to that observed with MPICH-dev.
Comparing Figure~\ref{mpich43} and Figure~\ref{mpich-dev}, MPICH 4.3.0 demonstrates faster initialization times but significantly higher memory usage, especially at larger node counts.
The initialization time difference is mainly attributed to the redesign of the address exchange algorithm.
MPICH 4.3.0 assumes world initialization, whereas the new address exchange algorithm in MPICH-dev supports arbitrary communicator bootstrapping. This flexibility introduces additional steps for collecting and parsing process compositions, which increases the overall initialization time.
The memory consumption difference, on the other hand, is due to an unrelated optimization in the MPICH development branch. MPICH 4.3.0 statically allocates address tables for all potential internal communication endpoints. These tables are replicated for each network interface card (NIC) on the node and for every process. In contrast, MPICH-dev employs a more dynamic approach to address table allocation, significantly reducing memory usage at the cost of a negligible increase in runtime latency when additional address vector (AV) entries are required for new contexts.

Running Open MPI on Aurora, especially with the session model, presents several challenges. We experienced crashes or hangs using the system job launcher (PALS from the Cray Programming Environment), in particular when performing group-level communication bootstrap. Thus, we ran experiments using Open MPI's \texttt{mpiexec} instead. Even so, we encountered instability at high node counts, and were unable to obtain results at 1024 and 2048 nodes due to frequent job failures.

\begin{figure} 
    \centering
\resizebox{0.95\textwidth}{!}{
    \input{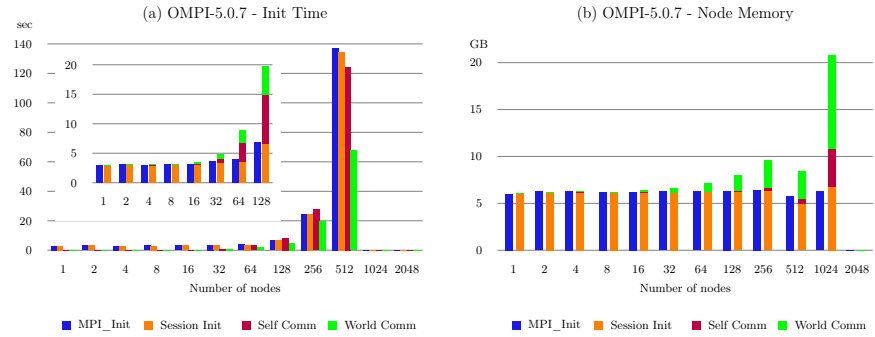}
}    
\caption{Comparing MPI Initialization between the world model and the session model using Open MPI 5.0.7. (a) Initialization times in seconds. A zoomed section show the same data from 1 to 128 nodes. (b) Node memory usage in GB.}
\label{ompi5}
\end{figure}

As shown in Figure~\ref{ompi5}, Open MPI uses less memory overall but exhibits \linebreak longer initialization time compared to MPICH. Notably, the time for both \linebreak \SessionInit\ and self communicator bootstrapping increases super-linearly with the number of nodes. Although neither step is collective, we suspect that some components of the software stack may still introduce implicit synchronization overhead, contributing to the observed scaling inefficiencies.
Using Open MPI's bundled \texttt{mpiexec} on Aurora is likely to have integration issues with the system resource manager. Thus, further insights are needed to interpret this experiment.

\subsection{Sparse World Initialization}
By supporting true \sessions, applications can bypass the creation of a global world communicator altogether. One compelling use case is a sparsely connected hierarchical topology, in which all processes on a node form a node communicator, and a designated node root from each node joins a second communicator, the node roots communicator.
This hierarchical structure still enables global collective operations through explicitly programmed algorithms. For example, an \texttt{MPI\_Allreduce} can be implemented as follows:
\begin{lstlisting}
  MPI_Reduce(buf, recvbuf, count, datatype, MPI_SUM, 0, node_comm);
  if (node_comm->rank == 0) {
    MPI_Allreduce(MPI_IN_PLACE, recvbuf, count, datatype, MPI_SUM, node_roots_comm);
  }
  MPI_Bcast(recvbuf, count, datatype, 0, node_comm);
\end{lstlisting}

Let $m$ represent the number of processes per node (PPN). Compared to a fully connected world communicator, this hierarchical setup reduces the number of internode connections by approximately a factor of $m^2$. The actual resource savings, however, depends on the overhead associated with establishing internode connections.

% The session measurements are split into MPI\_Session\_init, creating node communicator, and creating node roots communicator (only on node root process).

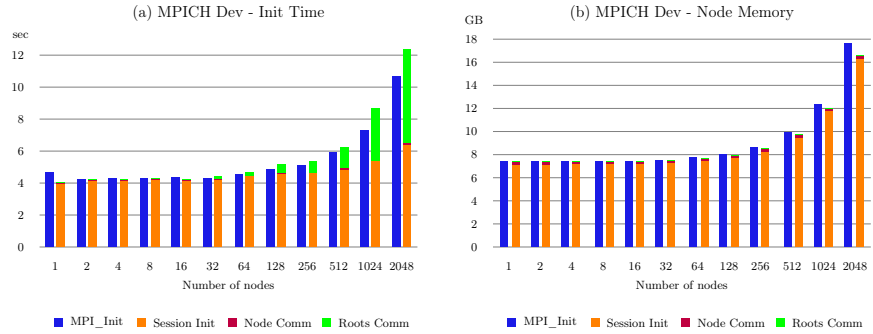
\begin{figure}
    \centering
\resizebox{0.95\textwidth}{!}{
    \begin{tikzpicture}[
    dummy/.style={fill=blue!80!gray},
    sess_init/.style={fill=orange},
    sess_self/.style={fill=purple},
    sess_world/.style={fill=green},
]
% title
\node at (5, 6.2) {\large (a) MPICH Dev - Init Time};
% x labels
\draw (0.4167, 0-0.5) node {1};
\draw (1.25, 0-0.5) node {2};
\draw (2.083, 0-0.5) node {4};
\draw (2.917, 0-0.5) node {8};
\draw (3.75, 0-0.5) node {16};
\draw (4.583, 0-0.5) node {32};
\draw (5.417, 0-0.5) node {64};
\draw (6.25, 0-0.5) node {128};
\draw (7.083, 0-0.5) node {256};
\draw (7.917, 0-0.5) node {512};
\draw (8.75, 0-0.5) node {1024};
\draw (9.583, 0-0.5) node {2048};
% y labels
\draw (0-0.3, 0) node[anchor=east] {0};
\draw[gray] (0, 0) -- (10, 0);
\draw (0-0.3, 0.8462) node[anchor=east] {2};
\draw[gray] (0, 0.8462) -- (10, 0.8462);
\draw (0-0.3, 1.692) node[anchor=east] {4};
\draw[gray] (0, 1.692) -- (10, 1.692);
\draw (0-0.3, 2.538) node[anchor=east] {6};
\draw[gray] (0, 2.538) -- (10, 2.538);
\draw (0-0.3, 3.385) node[anchor=east] {8};
\draw[gray] (0, 3.385) -- (10, 3.385);
\draw (0-0.3, 4.231) node[anchor=east] {10};
\draw[gray] (0, 4.231) -- (10, 4.231);
\draw (0-0.3, 5.077) node[anchor=east] {12};
\draw[gray] (0, 5.077) -- (10, 5.077);
% legend
\node at (-0.5, 5.6) {sec};
\node at (5, -1) {Number of nodes};
\node[dummy, minimum width=0.3, minimum height=0.3] at (0.5, -2) {};
\node[anchor=west] at (0.5+0.2, -2) {MPI\_Init};
\node[sess_init, minimum width=0.3, minimum height=0.3] at (2.7, -2) {};
\node[anchor=west] at (2.7+0.2, -2) {Session Init};
\node[sess_self, minimum width=0.3, minimum height=0.3] at (5.1, -2) {};
\node[anchor=west] at (5.1+0.2, -2) {Node Comm};
\node[sess_world, minimum width=0.3, minimum height=0.3] at (7.6, -2) {};
\node[anchor=west] at (7.6+0.2, -2) {Roots Comm};
% MPI_Init
\fill[dummy] (0.167, 0) rectangle (0.383, 1.98);
\fill[sess_init] (0.45, 0) rectangle (0.667, 1.67);
\fill[sess_self] (0.45, 1.67) rectangle (0.667, 1.7);
\fill[sess_world] (0.45, 1.7) rectangle (0.667, 1.7);
\fill[dummy] (1, 0) rectangle (1.22, 1.79);
\fill[sess_init] (1.28, 0) rectangle (1.5, 1.76);
\fill[sess_self] (1.28, 1.76) rectangle (1.5, 1.78);
\fill[sess_world] (1.28, 1.78) rectangle (1.5, 1.78);
\fill[dummy] (1.83, 0) rectangle (2.05, 1.81);
\fill[sess_init] (2.12, 0) rectangle (2.33, 1.76);
\fill[sess_self] (2.12, 1.76) rectangle (2.33, 1.79);
\fill[sess_world] (2.12, 1.79) rectangle (2.33, 1.79);
\fill[dummy] (2.67, 0) rectangle (2.88, 1.81);
\fill[sess_init] (2.95, 0) rectangle (3.17, 1.78);
\fill[sess_self] (2.95, 1.78) rectangle (3.17, 1.8);
\fill[sess_world] (2.95, 1.8) rectangle (3.17, 1.8);
\fill[dummy] (3.5, 0) rectangle (3.72, 1.84);
\fill[sess_init] (3.78, 0) rectangle (4, 1.76);
\fill[sess_self] (3.78, 1.76) rectangle (4, 1.79);
\fill[sess_world] (3.78, 1.79) rectangle (4, 1.79);
\fill[dummy] (4.33, 0) rectangle (4.55, 1.82);
\fill[sess_init] (4.62, 0) rectangle (4.83, 1.78);
\fill[sess_self] (4.62, 1.78) rectangle (4.83, 1.81);
\fill[sess_world] (4.62, 1.81) rectangle (4.83, 1.88);
\fill[dummy] (5.17, 0) rectangle (5.38, 1.92);
\fill[sess_init] (5.45, 0) rectangle (5.67, 1.87);
\fill[sess_self] (5.45, 1.87) rectangle (5.67, 1.89);
\fill[sess_world] (5.45, 1.89) rectangle (5.67, 1.99);
\fill[dummy] (6, 0) rectangle (6.22, 2.04);
\fill[sess_init] (6.28, 0) rectangle (6.5, 1.94);
\fill[sess_self] (6.28, 1.94) rectangle (6.5, 1.96);
\fill[sess_world] (6.28, 1.96) rectangle (6.5, 2.19);
\fill[dummy] (6.83, 0) rectangle (7.05, 2.16);
\fill[sess_init] (7.12, 0) rectangle (7.33, 1.96);
\fill[sess_self] (7.12, 1.96) rectangle (7.33, 1.98);
\fill[sess_world] (7.12, 1.98) rectangle (7.33, 2.28);
\fill[dummy] (7.67, 0) rectangle (7.88, 2.49);
\fill[sess_init] (7.95, 0) rectangle (8.17, 2.06);
\fill[sess_self] (7.95, 2.06) rectangle (8.17, 2.09);
\fill[sess_world] (7.95, 2.09) rectangle (8.17, 2.64);
\fill[dummy] (8.5, 0) rectangle (8.72, 3.09);
\fill[sess_init] (8.78, 0) rectangle (9, 2.27);
\fill[sess_self] (8.78, 2.27) rectangle (9, 2.29);
\fill[sess_world] (8.78, 2.29) rectangle (9, 3.67);
\fill[dummy] (9.33, 0) rectangle (9.55, 4.52);
\fill[sess_init] (9.62, 0) rectangle (9.83, 2.72);
\fill[sess_self] (9.62, 2.72) rectangle (9.83, 2.75);
\fill[sess_world] (9.62, 2.75) rectangle (9.83, 5.22);
% title
\node at (17, 6.2) {\large (b) MPICH Dev - Node Memory};
% x labels
\draw (12.42, 0-0.5) node {1};
\draw (13.25, 0-0.5) node {2};
\draw (14.08, 0-0.5) node {4};
\draw (14.92, 0-0.5) node {8};
\draw (15.75, 0-0.5) node {16};
\draw (16.58, 0-0.5) node {32};
\draw (17.42, 0-0.5) node {64};
\draw (18.25, 0-0.5) node {128};
\draw (19.08, 0-0.5) node {256};
\draw (19.92, 0-0.5) node {512};
\draw (20.75, 0-0.5) node {1024};
\draw (21.58, 0-0.5) node {2048};
% y labels
\draw (12-0.3, 0) node[anchor=east] {0};
\draw[gray] (12, 0) -- (22, 0);
\draw (12-0.3, 0.6111) node[anchor=east] {2};
\draw[gray] (12, 0.6111) -- (22, 0.6111);
\draw (12-0.3, 1.222) node[anchor=east] {4};
\draw[gray] (12, 1.222) -- (22, 1.222);
\draw (12-0.3, 1.833) node[anchor=east] {6};
\draw[gray] (12, 1.833) -- (22, 1.833);
\draw (12-0.3, 2.444) node[anchor=east] {8};
\draw[gray] (12, 2.444) -- (22, 2.444);
\draw (12-0.3, 3.056) node[anchor=east] {10};
\draw[gray] (12, 3.056) -- (22, 3.056);
\draw (12-0.3, 3.667) node[anchor=east] {12};
\draw[gray] (12, 3.667) -- (22, 3.667);
\draw (12-0.3, 4.278) node[anchor=east] {14};
\draw[gray] (12, 4.278) -- (22, 4.278);
\draw (12-0.3, 4.889) node[anchor=east] {16};
\draw[gray] (12, 4.889) -- (22, 4.889);
\draw (12-0.3, 5.5) node[anchor=east] {18};
\draw[gray] (12, 5.5) -- (22, 5.5);
% legend
\node at (11.5, 6) {GB};
\node at (17, -1) {Number of nodes};
\node[dummy, minimum width=0.3, minimum height=0.3] at (12.5, -2) {};
\node[anchor=west] at (12.5+0.2, -2) {MPI\_Init};
\node[sess_init, minimum width=0.3, minimum height=0.3] at (14.7, -2) {};
\node[anchor=west] at (14.7+0.2, -2) {Session Init};
\node[sess_self, minimum width=0.3, minimum height=0.3] at (17.1, -2) {};
\node[anchor=west] at (17.1+0.2, -2) {Node Comm};
\node[sess_world, minimum width=0.3, minimum height=0.3] at (19.6, -2) {};
\node[anchor=west] at (19.6+0.2, -2) {Roots Comm};
% MPI_Init
\fill[dummy] (12.2, 0) rectangle (12.4, 2.27);
\fill[sess_init] (12.5, 0) rectangle (12.7, 2.19);
\fill[sess_self] (12.5, 2.19) rectangle (12.7, 2.27);
\fill[sess_world] (12.5, 2.27) rectangle (12.7, 2.27);
\fill[dummy] (13, 0) rectangle (13.2, 2.26);
\fill[sess_init] (13.3, 0) rectangle (13.5, 2.18);
\fill[sess_self] (13.3, 2.18) rectangle (13.5, 2.25);
\fill[sess_world] (13.3, 2.25) rectangle (13.5, 2.25);
\fill[dummy] (13.8, 0) rectangle (14, 2.26);
\fill[sess_init] (14.1, 0) rectangle (14.3, 2.19);
\fill[sess_self] (14.1, 2.19) rectangle (14.3, 2.26);
\fill[sess_world] (14.1, 2.26) rectangle (14.3, 2.26);
\fill[dummy] (14.7, 0) rectangle (14.9, 2.27);
\fill[sess_init] (15, 0) rectangle (15.2, 2.19);
\fill[sess_self] (15, 2.19) rectangle (15.2, 2.26);
\fill[sess_world] (15, 2.26) rectangle (15.2, 2.26);
\fill[dummy] (15.5, 0) rectangle (15.7, 2.27);
\fill[sess_init] (15.8, 0) rectangle (16, 2.2);
\fill[sess_self] (15.8, 2.2) rectangle (16, 2.27);
\fill[sess_world] (15.8, 2.27) rectangle (16, 2.27);
\fill[dummy] (16.3, 0) rectangle (16.5, 2.3);
\fill[sess_init] (16.6, 0) rectangle (16.8, 2.23);
\fill[sess_self] (16.6, 2.23) rectangle (16.8, 2.3);
\fill[sess_world] (16.6, 2.3) rectangle (16.8, 2.3);
\fill[dummy] (17.2, 0) rectangle (17.4, 2.36);
\fill[sess_init] (17.5, 0) rectangle (17.7, 2.27);
\fill[sess_self] (17.5, 2.27) rectangle (17.7, 2.34);
\fill[sess_world] (17.5, 2.34) rectangle (17.7, 2.34);
\fill[dummy] (18, 0) rectangle (18.2, 2.45);
\fill[sess_init] (18.3, 0) rectangle (18.5, 2.36);
\fill[sess_self] (18.3, 2.36) rectangle (18.5, 2.43);
\fill[sess_world] (18.3, 2.43) rectangle (18.5, 2.43);
\fill[dummy] (18.8, 0) rectangle (19, 2.64);
\fill[sess_init] (19.1, 0) rectangle (19.3, 2.53);
\fill[sess_self] (19.1, 2.53) rectangle (19.3, 2.6);
\fill[sess_world] (19.1, 2.6) rectangle (19.3, 2.6);
\fill[dummy] (19.7, 0) rectangle (19.9, 3.04);
\fill[sess_init] (20, 0) rectangle (20.2, 2.89);
\fill[sess_self] (20, 2.89) rectangle (20.2, 2.96);
\fill[sess_world] (20, 2.96) rectangle (20.2, 2.97);
\fill[dummy] (20.5, 0) rectangle (20.7, 3.78);
\fill[sess_init] (20.8, 0) rectangle (21, 3.6);
\fill[sess_self] (20.8, 3.6) rectangle (21, 3.67);
\fill[sess_world] (20.8, 3.67) rectangle (21, 3.68);
\fill[dummy] (21.3, 0) rectangle (21.5, 5.38);
\fill[sess_init] (21.6, 0) rectangle (21.8, 5);
\fill[sess_self] (21.6, 5) rectangle (21.8, 5.08);
\fill[sess_world] (21.6, 5.08) rectangle (21.8, 5.08);
\end{tikzpicture}
}    
\caption{Comparing MPI Initialization between the world model and the Sessions-based sparse model using mpich-dev. (a) Initialization times in seconds. (b) Node memory usage in GB.}
 \label{sparse}
\end{figure}

Figure~\ref{sparse} compares initialization time and memory usage between the traditional world model and the Sessions-based sparse model. The results confirm that constructing a sparse world using \sessions\ reduces both initialization time and memory consumption relative to building a full world communicator in the Sessions model (Figure~\ref{mpich-dev}). However, on Aurora, the traditional world model still initializes slightly faster. The sparse model does provide modest memory savings.

A significant portion of both the initialization time and memory usage occurs during local initialization, incurred not by MPICH itself but by its lower-layer dependencies. We suspect that these costs are mostly implementation issues rather than the fundamental limit. Many dependency layers in an HPC system assume and optimize for a traditional world model.

\section{Conclusions}
\sessions\ was introduced into MPI to address scalability challenges on exascale systems. We recently completed a significant code restructuring in MPICH to support \sessions\ in alignment with its original design goal—eliminating dependence on \WORLD.
We evaluated MPI initialization scalability by comparing equivalent use cases between the traditional world model and the sessions model. As expected, when a fully connected world communicator is required, the sessions model does not provide a performance advantage over the world model.
However, with our recent development, MPICH now supports the construction of sparsely connected communication structures without relying on a world communicator. Our experiments demonstrate that such sparse topologies can reduce memory usage and improve initialization time compared to the traditional all-to-all world communicator.

While \sessions\ may not be essential for scalability alone, transitioning from the world model to the Sessions model introduces greater flexibility and dynamism at minimal cost and without significant performance penalties.
This flexibility may become increasingly valuable for emerging and non-traditional application workflows.

MPI Sessions is a major addition in the MPI 4.0 standard, though it has yet to see widespread adoption. With robust support now available in MPICH, we aim to promote broader use and stimulate further research into the capabilities and practical use cases of \sessions.

\begin{credits}
\subsubsection{\ackname}
This research was supported by the U.S. Department of Energy, Office of Science, under Contract DE-AC02-06CH11357.

\subsubsection{\discintname}
The authors have no competing interests to declare that are relevant to the content of this article.
\end{credits}

\bibliographystyle{splncs04}
\bibliography{references}

@manual{mpi40,
    author = "{Message Passing Interface Forum}",
    title  = "{MPI}: A Message-Passing Interface Standard Version 4.0",
    url    = "https://www.mpi-forum.org/docs/mpi-4.0/mpi40-report.pdf",
    year   = 2021,
    month  = jun
}

@inproceedings{holmes2016mpi,
  title={{MPI Sessions}: Leveraging runtime infrastructure to increase scalability of applications at exascale},
  author={Holmes, Daniel and Mohror, Kathryn and Grant, Ryan E and Skjellum, Anthony and Schulz, Martin and Bland, Wesley and Squyres, Jeffrey M},
  booktitle={Proceedings of the 23rd European MPI Users' Group Meeting},
  pages={121--129},
  year={2016}
}

@inproceedings{fecht2022emulation,
  title={An emulation layer for dynamic resources with {MPI} sessions},
  author={Fecht, Jan and Schreiber, Martin and Schulz, Martin and Pritchard, Howard and Holmes, Daniel J},
  booktitle={International Conference on High Performance Computing},
  pages={147--161},
  year={2022},
  organization={Springer}
}

@inproceedings{rocco2023fault,
  title={Fault awareness in the {MPI} 4.0 session model},
  author={Rocco, Roberto and Palermo, Gianluca and Gregori, Daniele},
  booktitle={Proceedings of the 20th ACM International Conference on Computing Frontiers},
  pages={189--192},
  year={2023}
}

@inproceedings{balaji2010pmi,
  title={{PMI}: A scalable parallel process-management interface for extreme-scale systems},
  author={Balaji, Pavan and Buntinas, Darius and Goodell, David and Gropp, William and Krishna, Jayesh and Lusk, Ewing and Thakur, Rajeev},
  booktitle={European MPI Users' Group Meeting},
  pages={31--41},
  year={2010},
  organization={Springer}
}

@inproceedings{castain2017pmix,
  title={{PMIx}: Process management for exascale environments},
  author={Castain, Ralph H and Solt, David and Hursey, Joshua and Bouteiller, Aurelien},
  booktitle={Proceedings of the 24th European MPI Users' Group Meeting},
  pages={1--10},
  year={2017}
}

@inproceedings{yoo2003slurm,
  title={Slurm: Simple {Linux} utility for resource management},
  author={Yoo, Andy B and Jette, Morris A and Grondona, Mark},
  booktitle={Workshop on job scheduling strategies for parallel processing},
  pages={44--60},
  year={2003},
  organization={Springer}
}

@inproceedings{ahn2014flux,
  title={Flux: A next-generation resource management framework for large {HPC} centers},
  author={Ahn, Dong H and Garlick, Jim and Grondona, Mark and Lipari, Don and Springmeyer, Becky and Schulz, Martin},
  booktitle={2014 43rd International Conference on Parallel Processing Workshops},
  pages={9--17},
  year={2014},
  organization={IEEE}
}

@article{dosanjh2021implementation,
  title={Implementation and evaluation of {MPI} 4.0 partitioned communication libraries},
  author={Dosanjh, Matthew GF and Worley, Andrew and Schafer, Derek and Soundararajan, Prema and Ghafoor, Sheikh and Skjellum, Anthony and Bangalore, Purushotham V and Grant, Ryan E},
  journal={Parallel Computing},
  volume={108},
  pages={102827},
  year={2021},
  publisher={Elsevier}
}

@article{gropp1996high,
  title={A high-performance, portable implementation of the {MPI Message Passing Interface Standard}},
  author={Gropp, William and Lusk, Ewing and Doss, Nathan and Skjellum, Anthony},
  journal={Parallel computing},
  volume={22},
  number={6},
  pages={789--828},
  year={1996},
  publisher={Elsevier}
}

@inproceedings{thakur1999data,
  title={Data sieving and collective {I/O} in {ROMIO}},
  author={Thakur, Rajeev and Gropp, William and Lusk, Ewing},
  booktitle={Proceedings. Frontiers' 99. Seventh Symposium on the Frontiers of Massively Parallel Computation},
  pages={182--189},
  year={1999},
  organization={IEEE}
}

@article{nemesis2007,
  title={Implementation and evaluation of shared-memory communication and synchronization operations in {MPICH2} using the {Nemesis} communication subsystem},
  author={Buntinas, Darius and Mercier, Guillaume and Gropp, William},
  journal={Parallel Computing},
  volume={33},
  number={9},
  pages={634--644},
  year={2007},
  publisher={Elsevier}
}

@inproceedings{balaji2009mpi,
  title={{MPI} on a Million Processors},
  author={Balaji, Pavan and Buntinas, Darius and Goodell, David and Gropp, William and Kumar, Sameer and Lusk, Ewing and Thakur, Rajeev and Tr{\"a}ff, Jesper Larsson},
  booktitle={Recent Advances in Parallel Virtual Machine and Message Passing Interface: 16th European PVM/MPI Users’ Group Meeting, Espoo, Finland, September 7-10, 2009. Proceedings 16},
  pages={20--30},
  year={2009},
  organization={Springer}
}

@inproceedings{raffenetti2018locality,
  title={Locality-aware {PMI} usage for efficient {MPI} startup},
  author={Raffenetti, Ken and Bayyapu, Neelima and Durnov, Dmitry and Takagi, Masamichi and Balaji, Pavan},
  booktitle={2018 IEEE 4th International Conference on Computer and Communications (ICCC)},
  pages={624--628},
  year={2018},
  organization={IEEE}
}

@article{zhou2024generating,
  title={Generating Bindings in {MPICH}},
  author={Zhou, Hui and Raffenetti, Ken and Bland, Wesley and Guo, Yanfei},
  journal={arXiv preprint arXiv:2401.16547},
  year={2024}
}

@article{guo2025preparing,
  title={Preparing {MPICH} for exascale},
  author={Guo, Yanfei and Raffenetti, Ken and Zhou, Hui and Balaji, Pavan and Si, Min and Amer, Abdelhalim and Iwasaki, Shintaro and Seo, Sangmin and Congiu, Giuseppe and Latham, Robert and others},
  journal={The International Journal of High Performance Computing Applications},
  year={2025},
  pages={10943420241311608},
  publisher={SAGE Publications Sage UK: London, England}
}

@inproceedings{atchley2023frontier,
  title={Frontier: Exploring Exascale},
  author={Atchley, Scott and Zimmer, Christopher and Lange, John and Bernholdt, David and Melesse Vergara, Veronica and Beck, Thomas and Brim, Michael and Budiardja, Reuben and Chandrasekaran, Sunita and Eisenbach, Markus and others},
  booktitle={Proceedings of the International Conference for High Performance Computing, Networking, Storage and Analysis},
  pages={1--16},
  year={2023}
}

@article{suarez2024deep,
  title={The {DEEP-SEA} project: A software stack for heterogeneous and modular supercomputers},
  author={Suarez, Estela and Eicker, Norbert and Hoppe, Hans-Christian},
  year={2024},
  journal={PARS-Mitteilungen: Vol. 36},
  publisher={Gesellschaft f{\"u}r Informatik eV, Fachgruppe PARS}
}

@article{wozniak2019mpi,
  title={{MPI} jobs within {MPI} jobs: A practical way of enabling task-level fault-tolerance in {HPC} workflows},
  author={Wozniak, Justin M and Dorier, Matthieu and Ross, Robert and Shu, Tong and Kurc, Tahsin and Tang, Li and Podhorszki, Norbert and Wolf, Matthew},
  journal={Future Generation Computer Systems},
  volume={101},
  pages={576--589},
  year={2019},
  publisher={Elsevier}
}

@manual{intelmpi,
    author = "{Intel Corporation}",
    title  = "Intel\textsuperscript{\textregistered} {MPI} Library",
    url    = "https://www.intel.com/content/www/us/en/developer/tools/oneapi/mpi-library.html",
    year   = 2025
}

@manual{craympi,
    author = "{Hewlett Packard Enterprise}",
    title  = "Cray {MPICH}",
    url    = "https://cpe.ext.hpe.com/docs/24.03/mpt/mpich/index.html",
    year   = 2024
}

@manual{psmpi,
    author = "{ParTec AG}",
    title  = "ParaStation {MPI}",
    url    = "https://github.com/ParaStation/psmpi",
    year   = 2025
}

@manual{mvapich,
    author = "{The Ohio State University}",
    title  = "{MVAPICH}",
    url    = "https://mvapich.cse.ohio-state.edu/",
    year   = 2025
}

@manual{aurora,
    author = "{Argonne Leadership Computing Facility}",
    title = "Aurora",
    url = "https://www.alcf.anl.gov/aurora",
    year = 2025
}

\end{document}